# Nonequilibrium Transport Characteristics of substances in a Rough Potential Field

Peng Wang[1], Yang Zhang[1], Peng-Juan Zhang[1], Jie Huo[1,2], Xu-Ming Wang[1,2,*]

1 School of Physics and Electronic-Electrical Engineering, Ningxia University, Yinchuan 750021, PR China

2 Ningxia Key Laboratory of Intelligent Sensing for Desert Information, Yinchuan 750021, PR China

*E-mail: wang_xm@126.com

**Abstract**: A Langevin equation is proposed to describe the transport of overdamped Brownian particles in a periodic rough potential and driven by an unbiased periodic force. The equation can be transformed into the Fokker-Planck equation by using the Kramers-Moyal expansions. The time-dependent solution of Fokker-Planck equation demonstrates different modes of the probability flow. These modes include creeping in a single direction, direction reversal and oscillating in both directions in the coordinate space. By varying the roughness and noise intensity, the flow can transform between the modes. The correlation between the noise and space indicates that the noise can maintain the oscillation of the modes and prolong the transient time to the steady state at which the flow tends to be zero.

## 1 Introduction

Transport is one of the basic activities for a life system, and is a matter of great concern. But yet, the characteristics and the corresponding mechanism for many substances in different tissues remains uncertain due to the fact that the movement of the substances in an organism is sharply different from those in a general physical system. The particles or substance vehicles in a life system are often active ones that have self-driving ability, so the transport process can exhibit a series of collective effects such as hysteresis and non-equilibrium phase transitions [1]. As a result, many physicists use some appropriate physical ideas and methods such as stochastic dynamics [2], molecular dynamics [3], and complex networks [4] to explore the transport laws and mechanisms of life systems that are far away from equilibrium.

  The transport in many processes of life system are often directional, such as the movement of nutrients from blood vessels to tissues, and the entering of the nutrients and some active particles into the cell via passing through the membranes, etc. In the case of ratchet potential, the periodic force of which the average value is zero can still produce directional motion [5]. Although the effects of temporal-spatial asymmetry, non-equilibrium fluctuations, and other features have been extensively studied [5-7], the role of roughness in periodic potentials has been rarely explored. It is known, the spatial inhomogeneity or impurities can lead to deviation from the smooth ratchet profile, which will affect the transport efficiency, for instance, the roughness of the electrical potential is crucial in a biophysical environment [7-8]. Many observations support the same conclusion. The effect of the roughness in potential field has been experimentally measured in proteins [9]. The effective diffusion coefficient is reduced greatly compared to that on smooth surfaces [10]. The impurities or randomness can cause the disorder of the electric potential of the ratchet, and further lead to the quenching of thermal ratchet effectiveness [11]. The roughness significantly hinders the current flow for the overdamped Brownian dynamics in a one-dimensional periodic and rough ratchet potential [12].

In conclusion, the details of the transport caused by roughness remains unclever yet, the further investigations are expected. In our paper, the uncertainty of overdamped Brownian particles in a periodic rough potential and influenced by an unbiased periodic force is studied. By varying the roughness and noise intensity, we can observe a reversal of particle directional transport. This may be attributed to the cumulative effect caused by noise introduced via the coupling of noise with space.

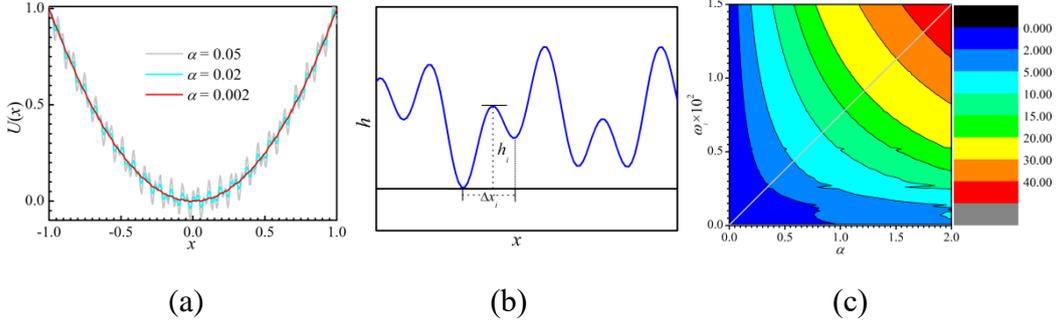

(a)  (b)  (c)

Fig. 1 (a) Spatial distribution of roughness potential field; (b) Local schematic diagram of detrended roughness potential field; (c) Phase diagram of roughness in parameter space.

## 2 Stochastic Dynamics Model for Transport

To explore the transport characteristics of overdamped Brownian particles in a rough potential field under an external driving force, the Langevin equation describing the dynamics is suggested as follows.

$$\gamma \frac{dx}{dt} = -\frac{\partial U(x)}{\partial x} + F(t) + \xi(t) \tag{1}$$

where $x$ is the position of the Brownian particle, $\gamma$ the damping coefficient, $F(t) = A\delta(t)$ is the external driving force with magnitude $A$, $\xi(t)$ white Gaussian noise, which satisfies the statistical relationship,

$$\begin{cases} \langle \xi(t) \rangle = 0 \\ \langle \xi(t)\xi(t') \rangle = 2D\delta(t-t') \end{cases} \tag{2}$$

$U(x)$ describes the rough potential field in Eq. (1), and is defined as

$$U(x) = x^2 + \alpha(\cos(\omega_1 x) + \sin(\omega_2 x)) \tag{3}$$

where $x^2$ is the classical harmonic potential, the rough potential field is shown by Fig.1 (a). Here we introduce such a rough potential to try to understand the oscillatory transport phenomenon in the organism. Naturally, the transport depends on the combination of parameter $\alpha, \omega_1, \omega_2$. To simplify the parameter description of the rough potential field, we reference the quantification method for the surface roughness proposed by Linden et al. [13]. As shown in Fig.1(b), a new parameter, the ratio of the net height $h_i$ of the rough surface to the horizontal distance $\Delta x_i$, is introduce. So the slope of the rough surface, and then the root mean square roughness is defined as

$$Ra = \left( \frac{1}{L} \sum_{i=1}^{n} \left( \frac{h_i}{\Delta x_i} - \overline{\left(\frac{h_i}{\Delta x_i}\right)} \right)^2 \right)^{\frac{1}{2}} \tag{3}$$

where $L$ is the spatial length of the measured rough surface, and $\overline{\left(\frac{h_i}{\Delta x_i}\right)}$ is the statistical average of the rough surface slope. The phase diagram of roughness in parameter space $(\alpha, \omega_1)$ is shown in Fig. 1(c). In region $\omega_1 < 20$, there is a local irregularity between roughness and parameters, while in region $\omega_1 \geq 20$, the roughness is positively related to parameter $\alpha$, which indicates that the roughness can be increased by increasing any one of the above parameters. In the following text, we adjust the roughness by changing any two of the parameters.

By using Kramers-Moyal series expansion method, Eq. (1) can be transformed into the Fokker-Planck equation, the differential equation of probability $P(x,t)$,

$$\frac{\partial P(x,t)}{\partial t} = -\frac{\partial}{\partial x} D^{(1)}(x,t) P(x,t) + \frac{\partial^2}{\partial x^2} D^{(2)}(x,t) P(x,t) \tag{4}$$

On the right side of equation, the first term is the drift one, and the second is the diffusion one. where $D^{(n)}(x,t)$ represent the $n$-order Kramers-Moyal expansion coefficient, which can be defined as

$$D^{(n)}(x,t) = \frac{1}{n!} \lim_{\tau \to 0} \frac{\langle (x(t+\tau) - x(t))^n \rangle}{\tau} \tag{5}$$

According to this definition, the coefficients in Eq. (3) can be obtained as

$$\begin{cases} D^{(1)}(x,t) = -\nabla U(x) + F(t) \\ D^{(1)}(x,t) = D \\ D^{(n)}(x,t) = 0, \quad n \geq 3 \end{cases} \tag{6}$$

Substituting Eq. (6) into Eq. (4), we have

$$\frac{\partial P(x,t)}{\partial t} = L_{FP}(x,t) P(x,t) \tag{7}$$

Here, the differential operator $L_{FP}(x,t)$ can be divided into two parts, the time-dependent and time-independent terms. They can be written as

$$\begin{cases} L_{FP}(x,t) = L_1(x) + L_2(x,t) \\ L_1(x) = -\frac{\partial}{\partial x}(-\nabla U(x)) + D\frac{\partial^2}{\partial x^2} \\ L_2(x,t) = F(t)\frac{\partial}{\partial x} \end{cases} \tag{8}$$

Supposing there exists a stable solution, $P_{st}(x)$, it satisfies the following equation,

$$L_{FP}(x,t) P_{st}(x) = L_1(x) P_{st}(x) = 0 \tag{9}$$

The solution can be given as

$$P_{st}(x) = Ne^{-\frac{U(x)}{D}} \tag{10}$$

where $N$ is the normalization constant. According to Eq. (8), The solution of Eq. (7) can be decomposed into a stable part and a time-dependent one, that is, the solution of Eq. (7) can be written as,

$$P(x,t) = P_{st}(x) + P_{ext}(x,t) \tag{11}$$

So, we have

$$\begin{aligned}\dot{P}(x,t) &= (L_1(x) + L_2(x,t))(P_{st}(x) + P_{ext}(x,t)) \\ &= L_1(x)P_{ext}(x,t) + L_2(x,t)P_{st}(x) + L_2(x,t)P_{ext}(x,t)\end{aligned} \tag{12}$$

It is known that the time-dependent solution is essentially a deviation from the stable solution, so it can be considered as a small quantity. And then the higher-order term of the time interval can be ignored, Eq. (12) can retain only the linear term,

$$\dot{P}_{ext}(x,t) = L_1(x)P_{ext}(x,t) + L_2(x,t)P_{st}(x) \tag{13}$$

The formal solution of Eq. (13) can be written as

$$P_{ext}(x,t) = \int_{-\infty}^{t} e^{L_1(x)\cdot(t-t')} L_2(x,t') P_{st}(x) dt' \tag{14}$$

Substituting Eq. (8) and Eq. (10) into Eq.(14), we have

$$P_{ext}(x,t) = ANe^{\left(\frac{\partial}{\partial x}(\nabla U(x)) + D\frac{\partial^2}{\partial x^2}\right)t} \frac{\partial}{\partial x} e^{-\frac{U(x)}{D}} \tag{15}$$

Therefore, the solution of Eq. (7) can be written as

$$\begin{aligned}P(x,t) &= P_{st}(x) + P_{ext}(x,t) \\ &= N\left(1 - A\frac{\nabla U(x)}{D} e^{-\lambda(x)\cdot t}\right) e^{-\frac{U(x)}{D}},\end{aligned} \tag{16}$$

$$\lambda(x) = \frac{2 - \alpha(\omega_1^2 \cos(\omega_1 x) + \omega_2^2 \sin(\omega_2 x))}{D} - \frac{\alpha(\omega_1^3 \sin(\omega_1 x) - \omega_2^3 \cos(\omega_2 x))}{2x + \alpha(-\omega_1 \sin(\omega_1 x) + \omega_2 \cos(\omega_2 x))}$$

According to the definition of probability flow, it can be given as

$$\frac{\partial P(x,t)}{\partial t} = -\frac{\partial J(x,t)}{\partial x} \tag{17}$$

Then, the probability flow can be written as:

$$\begin{aligned}J(x,t) &= D^{(1)}(x,t)P(x,t) - \frac{\partial}{\partial x} D^{(2)}(x,t)P(x,t) \\ &= -\nabla U(x)P(x,t) - D\frac{\partial P(x,t)}{\partial x}\end{aligned} \tag{18}$$

## 3 Result Analysis

Fig. 2 presents the theoretical results, the variation of probability under certain parameters $\alpha = 1.0 \times 10^{-9}$ and $D = 6.5 \times 10^{-2}$, and Fig. 2(a) shows a gradual transform from an asymmetric Poisson distribution to a symmetric Gaussian distribution. Fig. 2(b) depicts the skewness (third-order central moment) that gradually tends to zero from a negative value. This indicates that the change process of the spatial distribution is from a left-biased Poisson distribution to a Gaussian distribution. As well know, the asymmetry of the distribution is the source of transport, which can enhance the mass transport.

In order to discuss the detailed characteristics of the substance distribution changing from the asymmetric one to the symmetric one, the corresponding probability flow is calculated by changing the noise intensity and roughness, respectively. The results are shown in Figs. 3 and 4.

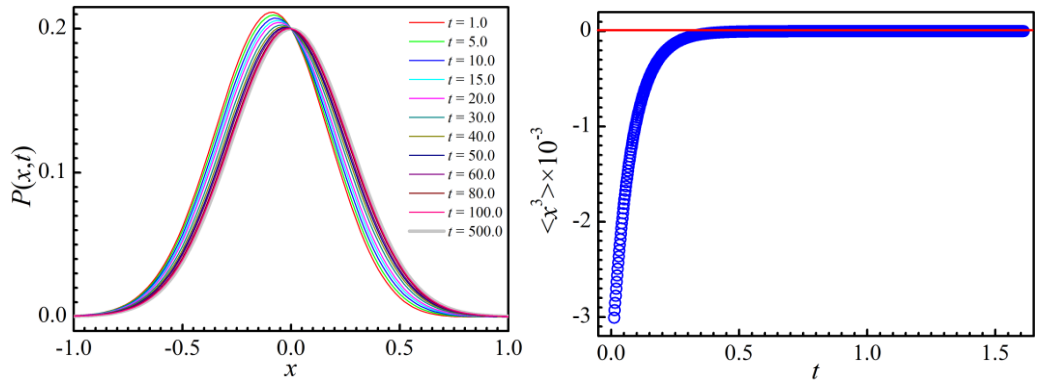

Fig. 2 (a) Variation of substance distribution; (b) Variation of skewness with time.

As the noise intensity is fixed to $D = 6.5 \times 10^{-3}$ (Fig. 3(a)), the probability flow at the initial time, $t = 1.0$, demonstrates an oscillating mode alternating positively and negatively with similar amplitude. With the passage of time ($t = 5.0, 10.0$), this oscillating flow decays and tends to zero flow, which indicates that the system reaches a stable distribution status.

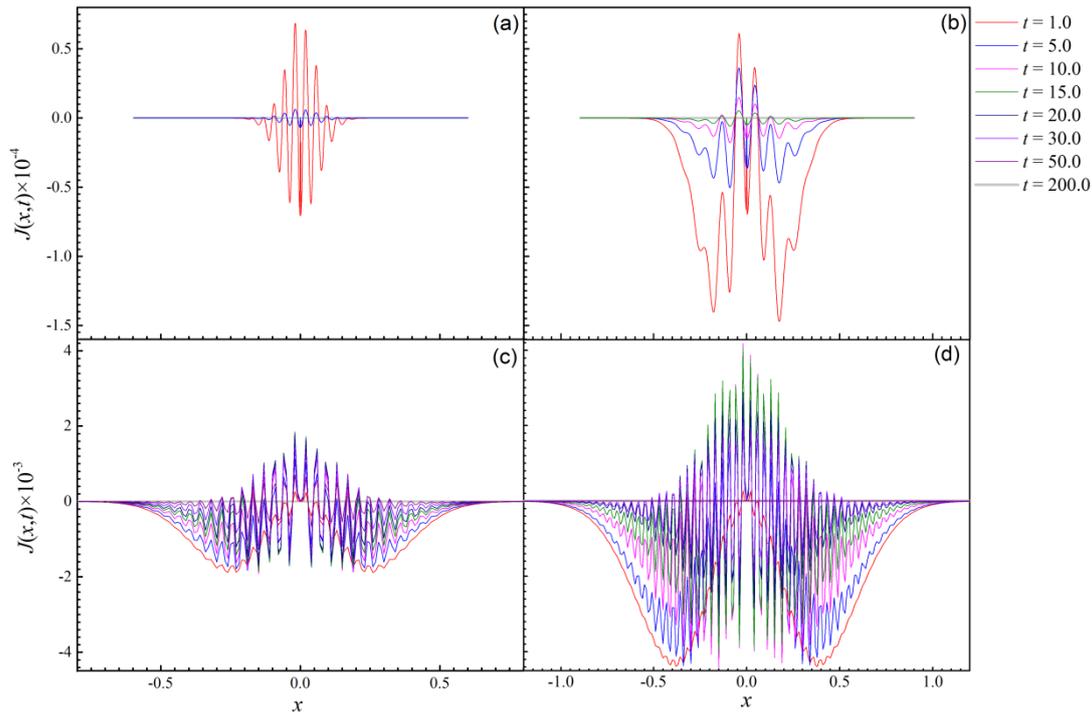

Fig. 3 Spatial distribution of probability flow under different noise intensities: (a) $D = 6.5 \times 10^{-3}$; (b) $D = 1.5 \times 10^{-2}$; (c) $D = 6.5 \times 10^{-2}$; (d) $D = 1.5 \times 10^{-1}$ 。

As the noise intensity increases, for instance, $D = 1.5 \times 10^{-2}$ (Fig. 3(b)), except for the small area near to the center $x = 0$ in which the distribution is dominated by the so-called positive-negative oscillating(PNO) mode shown in Fig. 3(a), the other region shows an negative(in the negative direction) irregular oscillating(NIO)mode. As time goes by, both the amplitude of the NIO mode and that of the PNO mode gradually weakens. From $t = 20$, these two modes disappear and the system tends to be at a stable status with zero flow. The NIO mode can be regarded as a creeping behavior of directional transport similar to the movement of a crawl insect, while the PNO mode implies that the positive-negative oscillation has the transport efficiency weakened. As the noise intensity is further enhanced, for instance, $D = 6.5 \times 10^{-2}$ (Fig. 3(c)), the amplitude of the flow oscillation increases and the center region, in which the PNO mode dominating the transport, becomes wider. By comparing Fig. 3(c) with Fig. 3(b), a change, flow reversal, occurs in the very small region near to and including the center $x = 0$ --the NIO mode become the PIO (positive irregular oscillating) mode. Finally, the flow in all modes decay to zero.

Fig. 3 (d) presents the results as the noise intensity increases to $D = 1.5 \times 10^{-1}$, the region of PNO mode becomes wider, and the amplitudes of the flow in all modes increase, and the flow gradually tends to zero.

The calculations above show that the transport always makes the system finally terminate in the establishment of a stable state of zero flow. Even more interesting is that, as shown by Fig. 4, the transient time before establishment of the stable state depends on the noise intensity in the manner of power law. From the discussions above, it is not difficult to find that the noise intensity causes the coexistence of and/or transition between modes of the PIO, NIO and PNO. However, what does this transition of flow modes mean in biological systems?

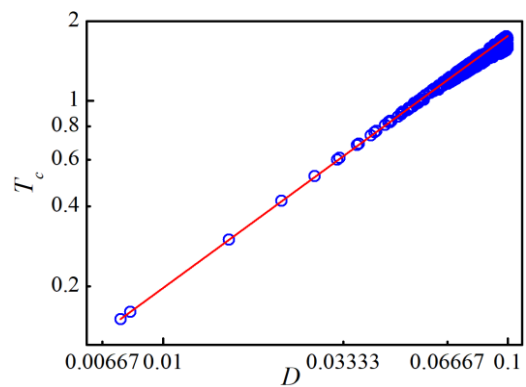

Fig. 4 The power-law relationship between the transient time before the steady state and the noise intensity.

To further explore the factors that cause the transition between the PIO, NIO and PNO modes dominating the transport, according to the phase diagram of the roughness parameter defined above, parameter $\omega_1$ is adjusted to change the roughness, and then the influence of the roughness on the flow is investigated. The results are shown in Fig. 5.

When the roughness is small ($\omega_1 = 36.7$), the spatial distribution of the probability flow is a smooth double-valley structure. The spatial distribution of the flow near $x < -1.0$, $x > 1.0$ and $x = 0.0$ tends to zero, and the negative flow of substance reaches a maximum value at the position of $x = \pm 0.4$. As shown in Fig. 5(a), the negative flow gradually decreases, and decays to zero at $t = 500.0$. This implies the system reaches a symmetrical Gaussian distribution. Increasing the roughness ($\omega_1 = 66.7$), the spatiotemporal evolution of the probability flow is shown in Fig. 5(b), which is similar to the above-mentioned spatiotemporal evolution, but with two differences. One is the local oscillation of the negative flow; the other is occurrence of weak PNO flow in the region near to $x = 0$. The former represents the creeping transport in the negative direction, the latter denotes a setback in the transport process. We note that transition from the NIO mode to the PNO mode occurs in the region there exist no or weak external driving force from the environment. As shown in Figs. 5 (c) and (d), the increasing roughness can lead to the continuous expansion of the PNO region. The dependence of the scale of PNO region $\Delta x$ on parameter $\omega_1$ denoting the roughness is shown by Fig. 6. As the roughness increases, the width of PNO region also increases nonlinearly.

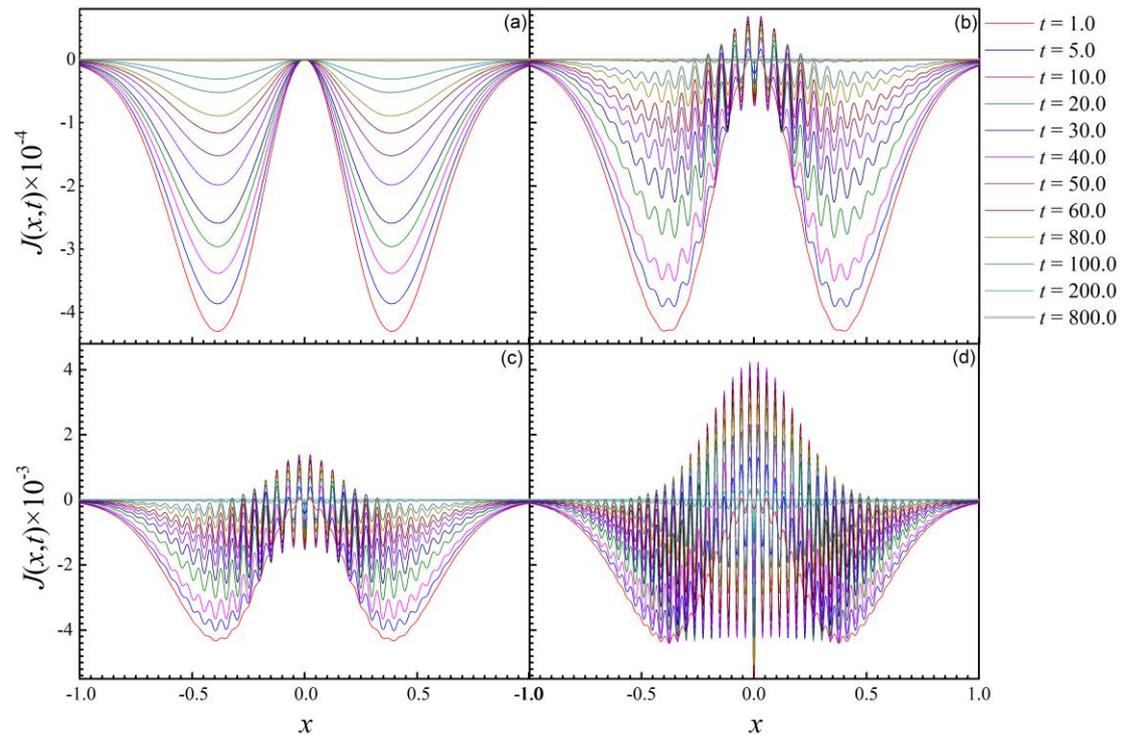

Fig. 5 Influence of the roughness on the flow: (a) $\omega_1 = 36.7$; (b) $\omega_1 = 86.7$;

(c) $\omega_1 = 126.7$; (d) $\omega_1 = 166.7$ 。

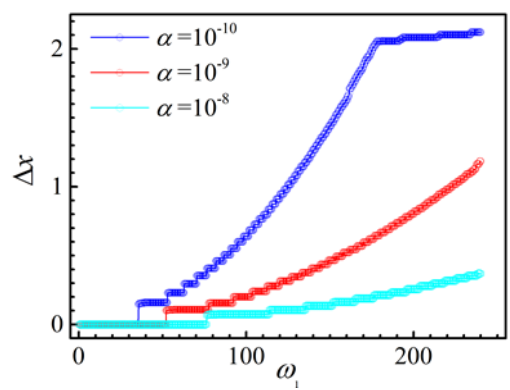

Fig. 6 The dependence of the parameter $\omega_1$, which controls the roughness, on the scale of PNO region $\Delta x$.

Based on the previous result that the transition of the transport mode from the creeping NIO to the PNO is caused by the increase of the roughness, so one may draw a conclusion that the roughness plays a role in counteracting the external field. We also note that there is no flow reversal near to the center in the roughness adjustment process, so one can infer from the discussion about influence of the noise intensity on the flow demonstrated by Figs. 3 and 4 that the most likely cause of the flow reversal in direction is the noise.

To verify the above inference, the correlation between noise and space is calculated, and is presented by Fig. 7. Obviously, there is a negative correlation between noise and space, and the correlation strength increases with the noise intensity. The negative correlation means that the correlation between noise and space intensifies the oscillation of both the NIO and PNO modes via increasing the amplitude, and inhibits the substance transport to some degree. This conclusion is in accordance with the results shown by Figs. 3 and 5. From another perspective, if one look at this negative correlation and what shown by Fig. 4 together, it is not difficult to draw a conclusion that the correlation between noise and space can suppress the attenuation of oscillation mode and prolong the transient time to the steady state.

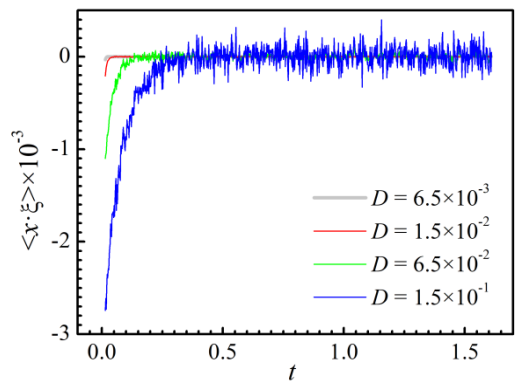

Fig. 7 Negative correlation effect of different noise and space.

## 4 Summary

This paper mainly studies the complex transport phenomenon of particles in rough potential. The results show that, under certain parameters, the spatial distribution of particles is gradually transformed from the initial asymmetric Poisson distribution to the symmetrical Gaussian distribution, and the critical time of this transformation process obeys a power-law relationship with the noise intensity. As the noise intensity varies, the probability flow can transform between the NIO, PIO, PNO modes. As the noise is weak, the flow is dominated by PNO mode and quickly decays to form a stable status; as the noise become stronger, the spatial region in divided in to two parts: in the central region the follow is governed by PNO mode, and in the other region the NIO mode drives the flow in the creeping way. And meanwhile the flow reversal occurs in the region near to the center to form PIO mode. As for as the influence of the roughness of the potential on the flow, in general, the rough potential is not conducive to directional transport. If the roughness is small, the negative flow forms the smooth double-valleys in space. As the roughness increases, the smooth double-valleys transform to the NIO mode to form a negative creeping flow. Meanwhile the flow will transform from the NIO to the PNO with time. The strong roughness can also increase the oscillating amplitude of the PNO and NIO modes and extend the region of the PNO. These variations indicate that the roughness can suppress the directional transport. The correlation between noise and space has a negative value and increases with the noise intensity. This implies that the correlation between noise and space suppresses the attenuation of the oscillation mode and prolongs the transient time to the steady state. The discussions may provide new understandings to the transport in living systems.

**Acknowledgements:** This study is supported by National Natural Science Foundation of China under Grant No. 11665018.

# References

[1] Tjhung E, Nardini C, Cates M E, Phys. Rev. X, 8: 031080(2018).

[2] Vasconcelos V V, Santos F P, Santos F C, Pacheco Jorge M, Phys. Rev. Lett., 118: 058301(2017).

[3] Das S K, Egorov S A, Trefz B, Virnau P, Binder K, Phys. Rev. Lett., 112: 198301(2014).

[4] Haimovici A, Tagliazucchi E, Balenzuela P, Chialvo D R, Phys. Rev. Lett., 110: 178101 (2013).

[5] P. Reimann, Phys. Rep. 361, 57 (2002).

[6] M. O. Magnasco, Phys. Rev. Lett. 71, 1477 (1993).

[7] R. Bartussek, P. Hanggi, and J. G. Kissner, Europhys. Lett. 28, 459 (1994).

[8] H. Frauenfelder, S. G. Sligar, and P. G. Wolynes, Science 254, 1598 (1991).

[9] M. V olk, L. Milanesi, J. P. Waltho, C. A. Hunter, and G. S. Beddard, Phys. Chem. Chem. Phys. 17, 762 (2015).

[10] R. Zwanzig, Proc. Natl. Acad. Sci. U.S.A. 85, 2029 (1988).

[11] F. Marchesoni, Phys. Rev. E 56, 2492 (1997).

[12] D. Mondal, P. K. Ghosh and D. S. Ray, J. Chem. Phys. 130, 074703 (2009).

[13] D. R. Linden, D. M. Van Doren Jr., Soil Sci. Sco. Am. J. 50, 1561(1986).